\documentclass{aastex63}

\usepackage[utf8]{inputenc}
\usepackage[titletoc, title]{appendix}
\usepackage{CJKutf8}
\usepackage{float}
\usepackage{blindtext}
\usepackage{gensymb}

\usepackage{amsmath}

\newcommand{\be}{\begin{equation}}
\newcommand{\ee}{\end{equation}}
\def\lta{\,\raise 0.3 ex\hbox{$ < $}\kern -0.75 em
 \lower 0.7 ex\hbox{$\sim$}\,}
\def\gta{\,\raise 0.3 ex\hbox{$ > $}\kern -0.75 em
 \lower 0.7 ex\hbox{$\sim$}\,}

\newcommand{\age}{t_{\rm age}}

\def \Physics {Department of Physics, University of Michigan, Ann Arbor, MI 48109, USA}
\def \Astronomy {Department of Astronomy, University of Michigan, Ann Arbor, MI 48109, USA}
\def \Caltech {Geological and Planetary Sciences, California Institute of Technology, Pasadena, CA 91125, USA}

\graphicspath{{Figures/}}

\begin{document}
\shorttitle{Capture of Interstellar Objects}
\shortauthors{Napier et al.}

\title{On the Fate of Interstellar Objects Captured by our Solar System}

\correspondingauthor{Kevin J. Napier}
\email{kjnapier@umich.edu}

\author[0000-0003-4827-5049]{Kevin J.~Napier}
\affiliation{\Physics}
\author[0000-0002-8167-1767]{Fred C.~Adams}
\affiliation{\Physics}
\affiliation{\Astronomy}
\author[0000-0002-7094-7908]{Konstantin Batygin}
\affiliation{\Caltech}

\begin{abstract}
With the recent discoveries of interstellar objects ‘Oumuamua and Borisov traversing the solar system, understanding the dynamics of interstellar objects is more pressing than ever. These detections have highlighted the possibility that captured interstellar material could be trapped in our solar system. The first step in rigorously investigating this question is to calculate a capture cross section for interstellar objects as a function of hyperbolic excess velocity, which can be convolved with any velocity dispersion to compute a capture rate \citep{Napier2021}. Although the cross section provides the first step toward calculating the mass of alien rocks residing in our solar system, we also need to know the lifetime of captured objects. We use an ensemble of N-body simulations to characterize a dynamical lifetime for captured interstellar objects and determines the fraction of surviving objects as a function of time (since capture). We also illuminate the primary effects driving their secular evolution. Finally, we use the resulting dynamical lifetime function to estimate the current inventory of captured interstellar material in the solar system. We find that capture from the field yields a steady state mass of only $\sim10^{-13} M_{\earth}$, whereas the mass remaining from capture events in the birth cluster is roughly $10^{-9}M_{\earth}$. 
\end{abstract} 

\keywords{Solar system (1528), Dynamical evolution (421), Small solar system bodies (1469), Kuiper belt (893), Oort cloud (1157)}

\accepted{22 September, 2021}
\submitjournal{The Planetary Science Journal}

\section{Introduction} 
\label{sec:intro}

The recent discoveries of the irregular body ‘Oumuamua \citep{meech2017} and the comet Borisov \citep{jewitt2019} have precipitated a boom in studies of interstellar objects. A pertinent open question of both scientific and general interest is whether our solar system currently harbors any material of extrinsic origin. Currently no objects present overwhelming evidence of having originated outside of our solar system \citep{morby2020}. Given the complex dynamical architecture of the outer solar system it is not straightforward to determine whether an object is of interstellar origin. A more severe difficulty is that we do not know the types of orbits on which such objects might reside. To address this issue, one must first determine the kinds of orbits onto which interstellar objects can be captured, and then determine which objects can survive for long enough to be considered quasi-permanent members of the solar system, rather than members of a transient population. 

The general process of gravitational capture, along with its application to our solar system, has been studied through both analytic and numerical treatments \citep{heggie1975, valtonen1982, valtonen1983, hands2020, Lehmann2021, Napier2021}. Our recent contribution (\citealt{Napier2021}; hereafter Paper I) used an ensemble of $\sim500$ million flyby numerical simulations to compute a capture cross section as a function of hyperbolic excess velocity. The resulting cross section, in conjunction with environmental parameters, allows one to compute a capture rate for any velocity distribution of interstellar objects. Although the specification of the capture rate is robust, we also need to know the dynamical lifetime of the captured objects in order to estimate the standing population of alien bodies. 

The final piece of this puzzle is thus to determine the dynamical stability and lifetimes for the population of captured interstellar objects. Previous related studies have investigated the dynamical stability of objects such as short-period centaurs \citep{RenuCentaurs}, short-period comets \citep{LevisonComets}, and long-period comets  \citep{MalyshkinComets}. However, these treatments cannot be directly applied to interstellar objects because the latter are captured on particular types of orbits -- with extremely high eccentricities, large semi-major axes, and perihelia lying in the giant planet regime \citep{Napier2021}. While several previous studies have now considered the capture of interstellar objects, none have offered a complete treatment of the long-term dynamics of the captured objects.

The objective of this paper is to examine the dynamical stability of captured interstellar objects and characterize their dynamical lifetime. We extend our previous work by performing long-term (up to 1 Gyr) simulations of 276,691 synthetic captured interstellar objects (Section 2). These results are then used to compute a dynamical lifetime function, i.e., the fraction of captured objects that remain in bound orbits as a function of time since capture (Section 3). We then consider (in Section 4) the secular evolution of the orbital elements of the captured objects in order to uncover the dynamics governing their survival at long times, and comment on the implications for our solar system.  Finally, we combine the dynamical lifetime function of this paper with the capture cross sections to obtain a refined estimate (in Section 5) of the total mass of alien rocks on bound orbits in the solar system (this estimate is limited to objects with pericenter within the orbit of Neptune). The paper concludes (in Section 6) with a summary of our results and a discussion of their implications.

\section{Numerical Simulations}

In order to understand and quantify the survival rate for captured interstellar objects, we use an ensemble of numerical simulations to study their long-term dynamical evolution and stability. More specifically, we perform numerical integrations for each of the 276,691 synthetic captured interstellar objects resulting from the simulations of Paper I. Note that these captured objects arise from encounters with given values of the initial (pre-capture) asymptotic speed $v_\infty$. We can thus determine how/if survival depends on $v_\infty$. 

The numerical simulation details are as follows. We use \texttt{Rebound's IAS15} integrator \citep{rebound} to track the orbit of each captured object for up to 1 Gyr (note that most of the objects are lost on shorter times scales). We use the same simulation details as Paper I, but instead of halting the simulation when an object becomes bound we continue the integration until one of the following events occurs: the object becomes unbound from the system, its apocenter distance exceeds the somewhat arbitrarily-chosen value of 50,000 au, it collides with a massive body, or the integration reaches 1 Gyr. Our simulations did not account for galactic tides and passing stars, as we assume that orbital diffusion is dominated by interactions with giant planets. We record the state vector of each captured object every 100 years for later use in studying the secular evolution of its orbit. 

Here we analyze the simulation data at the conceptual level to get a feel for the dynamics. We perform a more in-depth analysis of the long-lived objects in Section \ref{sec:long-lived}.
\begin{figure}[h]
    \centering
    \includegraphics[width=0.9\textwidth]{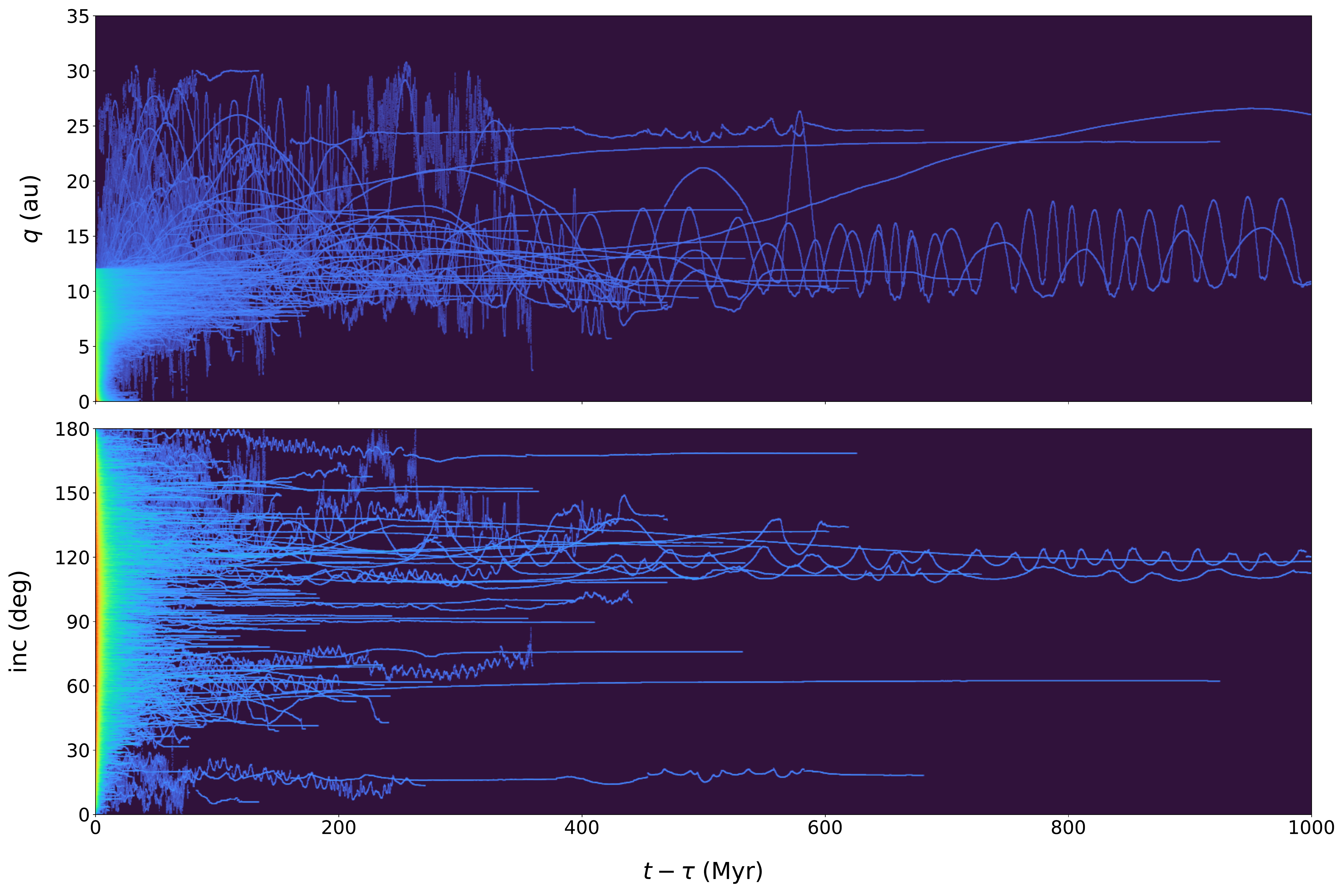}
    \caption{Time-series histogram of osculating pericenter distance (top) and inclination (bottom) for each of the objects we integrated. The color gradient corresponds to the point density. After approximately 100 Myr, individual objects become distinguishable.}
    \label{fig:time-series}
\end{figure}
In Figure \ref{fig:time-series} we show the time-series of each object's pericenter distance and inclination. Each line represents a single object, and truncates when the object was ejected from the system. It is difficult to make out any structure for the first $\sim$ 10 Myr, but soon thereafter the density of points falls off and some trends begin to emerge. There are a few features of note: 
\begin{itemize}
    \item[-] To survive for more than a few million years, captured objects must somehow lift their pericenters beyond Jupiter (survival in this context means that the object remains bound to the system).
    \item[-] Objects on highly inclined orbits tend to survive for longer than those on planar orbits.
    \item[-] No object achieved permanent trans-Neptunian status (i.e., $q > 30$ au).
\end{itemize}

These trends all make sense intuitively, and we briefly address each here. Objects that fail to lift their pericenters beyond Jupiter are likely to eventually have a close encounter with the planet and thus be scattered. Changes in pericenter can be achieved via mechanisms such as scattering or the von Ziepel-Lidov-Kozai (vZLK) effect \citep{vonZiepel, Lidov, kozai, Ito-vZLK}. Objects that are on highly inclined orbits are less likely to encounter a planet because they spend most of their time out of the plane of the solar system. However, objects on orbits nearly perpendicular to the plane of the solar system are subject to the polar instability, which can cause them to evolve onto unstable orbits and eventually be ejected from the system \citep{namouni}. Finally, in the absence of perturbations beyond Neptune, attaining a pericenter distance greater than 30 au requires some combination of serendipitous scattering events and vZLK resonances. 

For completeness, we note that our simulations make a number of minor approximations. First, the simulations include only the four giant planets and the Sun. We do not account for external forces such as out-gassing, radiation pressure during close approaches to the Sun, or drag due to orbits that cross through planetary atmospheres (which is an exceedingly rare event). In addition, we neglect the higher-order gravitational harmonics that arise from modeling the shape and rotation of massive bodies. Each of these approximations is rather modest, so that including them would make relatively little difference to our conclusions. We reiterate that our simulations did not model galactic or cluster tides, and did not include perturbations by passing stars. Incorporating these effects could allow some of the captured objects to lift their perihelia beyond Neptune, which would make them more likely to survive for the age of the solar system. These effects should be explored more thoroughly in future work.

\section{Dynamical Lifetime}

In this section we use the results of our simulations to characterize a dynamical lifetime for captured interstellar objects. Specifically we calculate the probability that any object still resides in the solar system after a given time $t$, and we designate this function as $f(t)$. 

\subsection{Numerical Results} 
\label{sec:numlifetime}

\begin{figure}[H]
    \centering
    \includegraphics[width=0.9\textwidth]{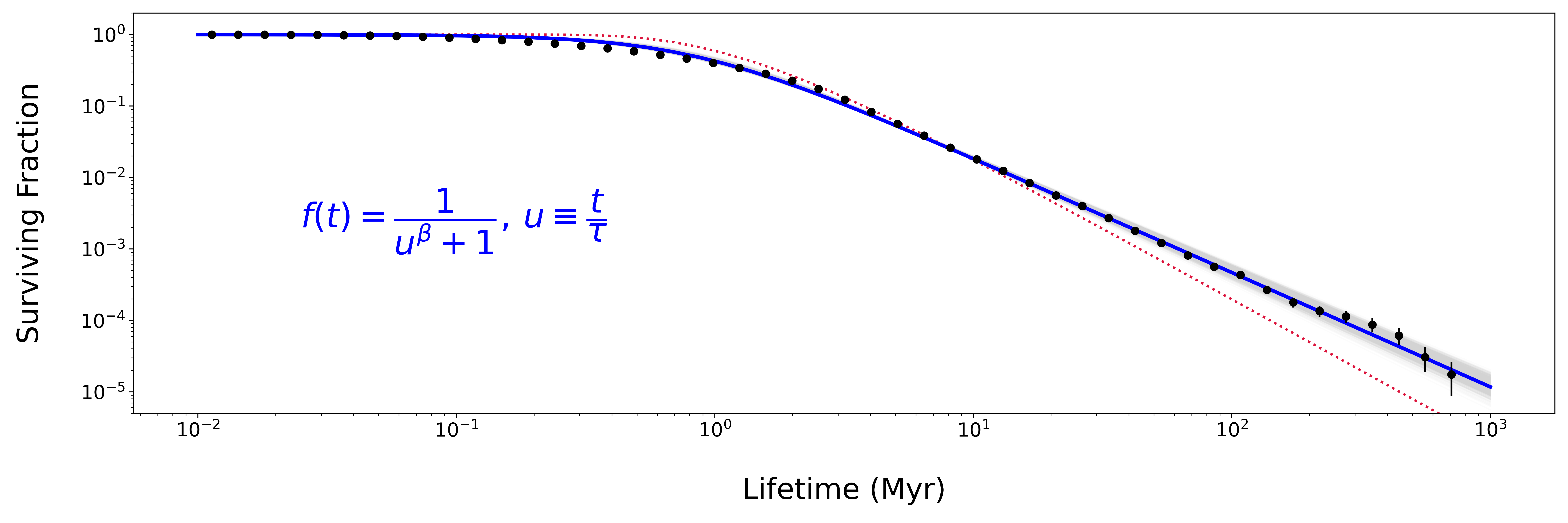}
    \caption{Surviving fraction of objects as a function of integration time. The black points represent the data, and the blue curve is the best fit using the expression of Equation (\ref{eq:lifetime}). The best-fit values are $\beta = 1.6 \pm 0.04$ and $\tau = 0.84 \pm 0.04$ Myr. The grey curve represents the uncertainty in the best fit. The dashed red line is an incomplete gamma function with index 2, as predicted by the diffusion model from \citet{yabushita}.}
    \label{fig:lifetime}
\end{figure}

The combined set of numerical data can be used to construct the fraction $f(t)$ of rocky bodies that survive as a function of time, where the result is shown in Figure \ref{fig:lifetime}. In addition, we can write down a simple function that fits the data, and find that the expression 
\begin{equation}
    f(t) = \frac{1}{u^\beta + 1} \qquad {\rm where} \qquad u \equiv \frac{t}{\tau}
    \label{eq:lifetime}
\end{equation}
agrees with our data over the entire range of lifetimes under consideration. Here the parameter $\tau$ acts as a characteristic timescale for the system, and $\beta$ is the power law index for depletion at long times. The best-fit values, with uncertainties calculated using a bootstrapping technique (see, e.g., \citealt{bootstrapping}), are $\tau \approx 0.84 \pm 0.04$ Myr, and $\beta \approx 1.6 \pm 0.04$ (note that we can take $\beta = 8/5$). In addition to our numerically determined survival function, Figure \ref{fig:lifetime} displays the best fit to Equation (\ref{eq:lifetime}) as the solid blue curve. The grey curve (shaded region) represents the uncertainty in the curve of best fit. For comparison, the figure also shows a previously obtained prediction for $f(t)$ from a diffusion model as the dotted red curve (see \citealt{yabushita} and Section \ref{sec:diffusion}). 

These data show that the population of rocky bodies remains relatively constant for a few million years before the system steadily ejects the captured objects. After this time, the lifetime function $f(t)$ takes the form of a power law. The fact that $f(t)$ transitions to a power law after a few million years has interesting implications. First we note that the orbits of captured object typically have semi-major axes $a\sim1000$ au with periods $P\sim30,000$ yr. It thus takes (at least) $\sim1$ Myr for enough orbits (here, more than 30) to accumulate to lead to ejection. In addition, the Lyapunov time of the outer solar system is roughly comparable to the transition time scale (e.g., \citealt{laskar1989}). This finding suggests that the system can be influenced by chaos over the time scales on which bodies are ejected. This result, in turn, affects theoretical descriptions of the ejection process (see Section \ref{sec:diffusion}).  The full theoretical interpretation is also complicated by effects such as secular resonances and vZLK resonances, which we discuss in Section \ref{sec:long-lived}. 

Although Equation (\ref{eq:lifetime}) fits the numerically determined lifetime function quite well, we note a few shortcomings. First, the sparse statistics in the number of long-lived objects (longer than, say, 100 Myr) makes the tail of the distribution somewhat uncertain. Although it is possible that the power law decay continues over even longer times than those explored here, it is also possible that the survival  fraction flattens out in the limit $t\to\infty$. In this case, we are underestimating the number of permanently captured objects. On the other hand, the large uncertainty for long lifetimes also allows the survival function to be somewhat steeper than the best fit power-law (with slope $\beta=8/5$). 

\begin{figure}
    \centering
    \includegraphics[width=0.95\textwidth]{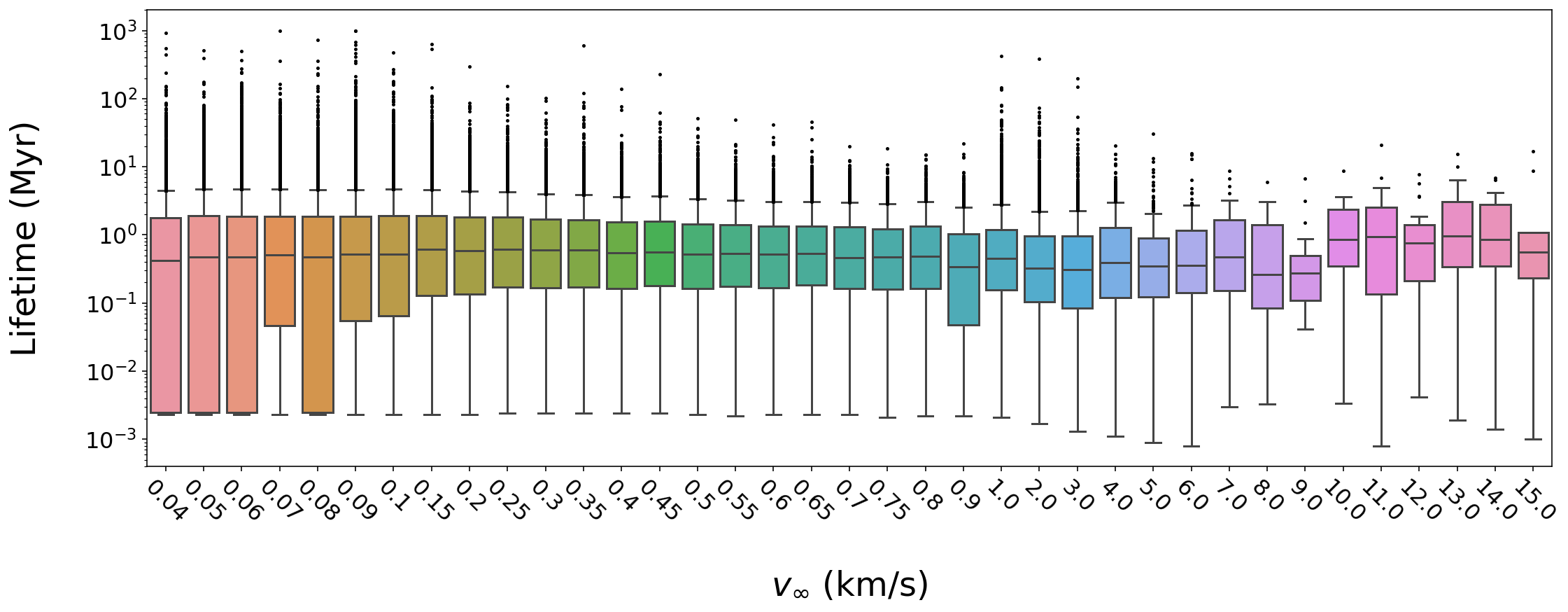}
    \caption{Box-and-whisker plots of dynamical lifetime as a function of hyperbolic excess velocity, $v_\infty$. There is not a noticeable correlation between $v_\infty$ and dynamical lifetime (and this vindicates our decision to simply integrate the objects from Paper I). Note that the data is not binned linearly in $v_\infty$.}
    \label{fig:vinf-lifetime}
\end{figure}

Next we note that the ensemble of objects considered here represents all of the objects captured in the numerical exploration of Paper I. As a result, this sample does not necessarily correspond to the starting conditions that would result from a particular capture scenario. On one hand, the starting ensemble of captured objects determines their initial distribution of energies, which affects their lifetimes. On the other hand, the velocity scale $v_\sigma$ in the cross section determines the typical speed of captured objects for bodies sourced from an initial velocity distribution with relatively high speeds (e.g., from the field).\footnote{The capture rate is determined by the integral $\langle v \sigma \rangle$. If we consider capture from the field, for example, the integrand is  proportional to $v_\sigma v^2\exp[-v^2/2s^2]$, where $s$ is the velocity dispersion. In the limit $s\gg v_\sigma$, the integrand peaks near $v=v_\sigma/\sqrt{3}$ and thus the integral has most of its support near $v\sim v_\sigma\sim0.42$ km/s.} In addition, we find that the survival time is largely independent of the speed $v_\infty$ that characterizes the initial (pre-capture) orbit of the incoming bodies. Figure \ref{fig:vinf-lifetime} shows the numerically determined dynamical lifetimes as a function of the initial hyperbolic velocity $v_\infty$. Although the velocity bins at extremely low speeds ($v_\infty \lta 0.1$ km/s) show a wider distribution of lifetimes, the distributions (and their means) are largely constant over the range of incoming speeds (this result is consistent with \citealt{Lehmann2021}).

\subsection{Diffusion Approach} 
\label{sec:diffusion} 

The simulations show that most captured objects survive for a time scale of order 1 Myr, and that the population steadily decreases at later times. It is useful to have a theoretical description of how long it takes interactions between the rocky bodies and the planets to increase the orbital energy to an unbound state. The surviving fraction depends on time according to $f\sim1/t^\beta$, with $\beta=8/5$, for times spanning $t$ = 1 Myr to 1 Gyr. Most of the captured objects have large semimajor axes, $a\sim10^3-10^4$ au, with periheila falling in the realm of the giant planets. The captured rocks spend most of their time at large distances from the Sun, but periodically enter the giant planet region where their orbits are perturbed. In the regime of interest --- long survival times --- the orbital periods are much shorter than the lifetimes, so that many perturbations (perihelion passages) are generally required for ejection. Moreover, given the four giant planets and the wide range of orbital elements $(a,e,i)$ for the captured bodies, the perturbations are largely uncorrelated. To leading order, one expects a random walk in the energy of the orbits, with the perihelion and inclination angle held nearly constant \citep{duncan1987,dones1996}. The rocks become unbound when the energy becomes positive.

While the literature does not contain a complete exploration of the dynamical lifetime of captured interstellar objects, previous works have attempted to characterize the dynamical lifetime of comets. In particular, \citet{yabushita} studied the survival of long-period comets under the approximation that comets undergo one-dimensional diffusion in energy, and found an analytic solution for $f(t)$ in the form of an incomplete gamma function. However, later papers (e.g., \citealt{duncan1987,dones1996,MalyshkinComets}) point out that the diffusion approximation (which assumes that the perturbations are truly random) cannot capture resonance sticking and related effects. In this spirit, \citet{MalyshkinComets} study the the dynamical lifetime using a Keplerian map, including a rather complicated interpolation function that describes the energy kicks experienced by a comet at perihelion as a function of the perturbing planet's phase (a \textit{kick function}). The $f(t)$ functions computed in this previous paper, while correct in context, are limited (for the present application) because they were computed using the three-body approximation with a narrow range of test particle inclinations. Due to the complexity of our (current) numerical simulations, which use all four giant planets and consider the full range of orbital inclination, the corresponding kick function would be considerably more complicated. This complexity, in turn, makes a random (diffusion) description more applicable. Although the challenge of constructing an analytic theory for the survival fraction $f(t)$ remains for future work, here we briefly describe results from the diffusion approximation. 

The dimensionless form of the diffusion equation can be written in the form 
\begin{equation}
{\partial n \over \partial t} = D 
{\partial^2 \over \partial x^2} \left( x^{3/2} n \right) \,, 
\label{diffuse} 
\end{equation}
where $x$ is the absolute value of the dimensionless energy (so that objects become unbound as $x\to0$). The factor of $x^{3/2}$ arises from Kepler's law, i.e., orbits with lower energy have larger semimajor axis $a=1/x$, longer periods $P\sim x^{-3/2}$, and thus diffuse more slowly. Note that with the orbital time accounted for, the dimensionless diffusion parameter $D$ is a constant (see \citealt{yabushita} for further discussion).\footnote{Note that the diffusion field $n$ appearing in Equation (\ref{diffuse}) represents the  distribution of energies. Equivalently, one can define the quantity $\nu(x,t) = x^{3/2} n(x,t)$, which represents the distribution of bodies passing through perihelion per unit time. The scaled function $\nu(x,t)$ obeys a diffusion equation with the $x^{3/2}$ factor appearing outside the spatial derivative operator (compare \citealt{MalyshkinComets} and \citealt{yabushita}). } \cite{yabushita} derives the Green's function for the diffusion equation (\ref{diffuse}) and finds the solution for $n(x,t)$ corresponding to an initial distribution of energies $\delta(x-x_0)$, where $x_0$ is the starting dimensionless energy for all of the orbits. The resulting solution for the total number of surviving bodies $N(t)$ becomes 
\begin{equation}
    N(t) = \int_0^\infty n(x,t) dx = 
    \gamma[ 2,4\sqrt{x_0}/(Dt) ] \rightarrow {8x_0 \over (Dt)^2}\,,
\end{equation}
where $\gamma(2,\xi)$ is an incomplete gamma function \citep{absteg1972}. The final expression shows the survival probability in the asymptotic limit $t\to\infty$. The full form of the function is shown as the red dotted curve in Figure \ref{fig:lifetime}. Although the asymptotic slope is somewhat steeper than the best fit to the numerical data, this straightforward result provides a good working description.\footnote{For completeness, we note that the inclusion of resonance sticking \citep{MalyshkinComets} predicts a survival function $f(t)\sim t^{-p}$, where the power-law index is smaller than the numerical results. Specifically, $p\approx2/3$ for the standard case and $p\approx4/3$ if orbits are considered to become unbound with a small but negative energy.} 


\section{Analysis of long-lived objects}
\label{sec:long-lived}

In this section we more closely analyze the secular dynamics of the captured objects in order to uncover the effects governing the long-term evolution of their orbits. \citet{MalyshkinComets} found that the tail of their lifetime function was influenced by comets being locked into mean motion resonance with a giant planet. However, as previously discussed, this work considered the three-body problem with low inclinations. In the present problem, the bodies under consideration are on planet-crossing orbits in the presence of all four giant planets and explore a wide range of orbital inclination. These complications make it more difficult for a body to be captured into or remain in a mean motion resonance. Though we did not explicitly test our simulations for mean motion resonance, we did not note any obvious signs when examining the orbital elements of long-lived objects.

\begin{figure}[h]
    \centering
    \includegraphics[width=0.8\textwidth]{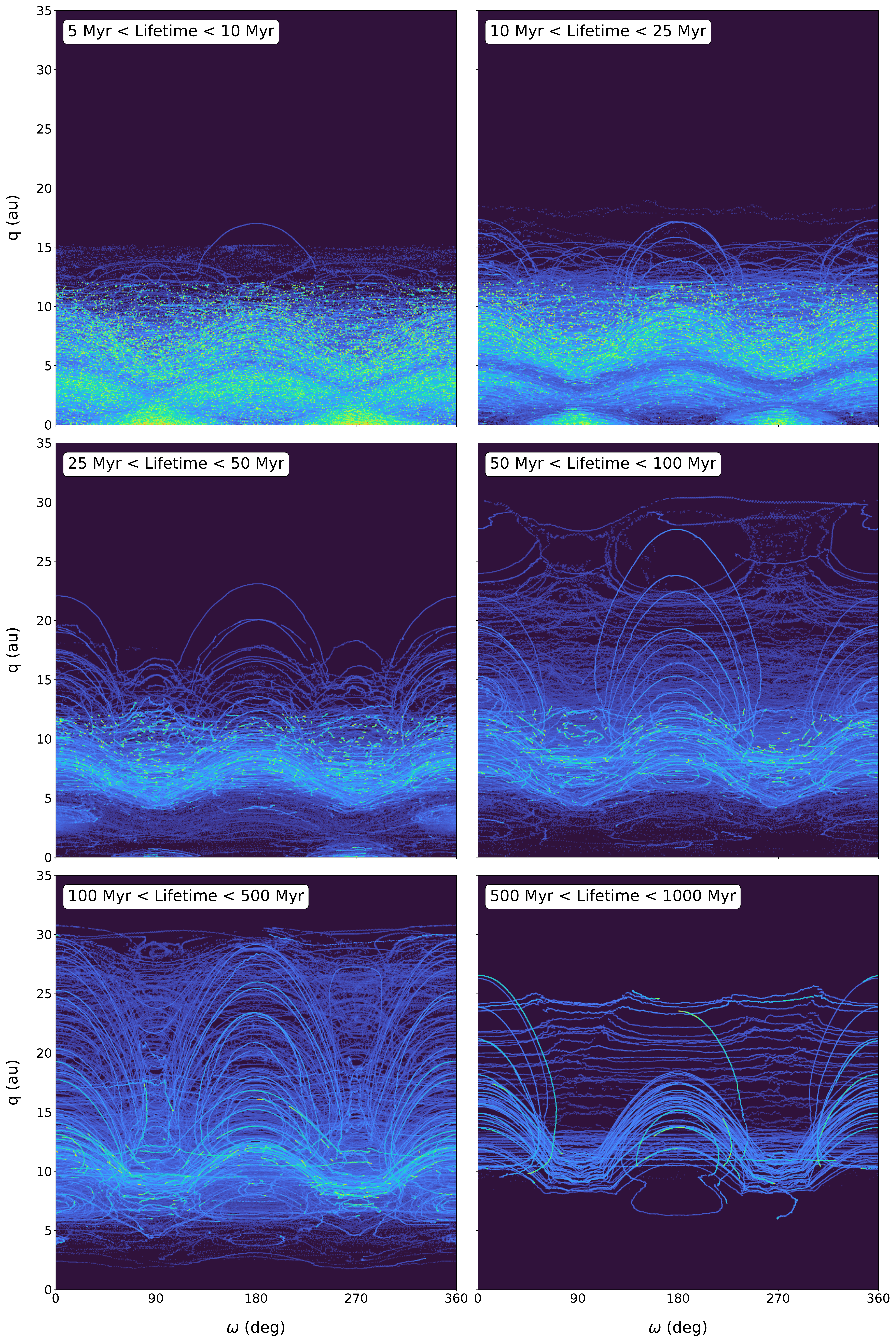}
    \caption{Two-dimensional histograms of $q$ vs $\omega$ for the captured objects, grouped into lifetime ranges. The color gradient corresponds to the point density. At long times, it becomes apparent that the captured objects are locked into vZLK oscillations (both librating and circulating).}
    \label{fig:kozai}
\end{figure}

A notable feature of our data is that none of the prograde captured objects survived the full Gyr-integration (although one object came close). Three retrograde objects survived the full simulation, and all three had similar inclinations ($i \sim 120\degree$). Though these statistics are not robust enough to make a definitive statement, we note an apparent trend in Figure \ref{fig:time-series}. The objects that can achieve and maintain inclinations $i\sim 60\degree$ with respect to the ecliptic (or invariable) plane are more likely to survive for the age of the solar system. This trend suggests that a common mechanism is driving the dynamical evolution of the longest-lived objects. Motivated by this finding, we investigate our data for signatures of vZLK resonances by plotting pericenter distance ($q$) vs argument of pericenter ($\omega$) in Figure \ref{fig:kozai}. We note that as dynamical lifetime increases, features resembling vZLK dynamics become more prominent. 


Given the numerical results, we now investigate the possible presence of vZLK dynamics analytically. The Hamiltonian for this system can be written as
\begin{equation}
    \mathcal{H} = \mathcal{H}_0 + \mathcal{P}
\end{equation}
where $\mathcal{H}_0$ is the unperturbed Keplerian part of the Hamiltonian, and $\mathcal{P}$ accounts for planetary perturbations. By limiting $\mathcal{P}$ to first order in planetary mass, one can average over the mean anomaly of both the test particle and planetary orbits. This averaging has the effect of fixing the semi-major axes of all bodies in the problem, meaning that we can simply drop the $\mathcal{H}_0$ term from the Hamiltonian, as it becomes a constant offset. Then, up to this constant offset, the Hamiltonian becomes the sum of doubly-averaged planetary perturbations, 
\begin{equation}
    \mathcal{K} = \frac{1}{4\pi^2}\sum_{k} m_k \int_0^{2\pi}\int_0^{2\pi} \mathcal{P}_k (M_k, a, e, i, \omega, \Omega, M) \, dM_k dM.
\end{equation}
In this expression the index $k$ runs through the giant planets, masses ($m_k$) are in units of solar masses, and distances are in units of au. By making the approximation that the planets are on circular orbits with zero inclination, $\mathcal{K}$ becomes rotationally invariant about $\Omega$ (for definiteness, we set $\Omega = 0$). \citet{Thomas1996} showed that after evaluating the integral with respect to $M_k$, the function takes the form 
\begin{equation}
    \mathcal{K} = \frac{1}{4\pi^2} \sum_{k} m_k \int_0^{2\pi} \frac{4}{\sqrt{r^2 + a_k^2}} \sqrt{1 - \frac{\xi}{2}} K(\xi) \, dM \,,
\end{equation}
where
\begin{equation}
    \xi = \frac{4 a_k \sqrt{x^2 + y^2}}{r^2 + a_k^2 + 2 a_k \sqrt{x^2 + y^2}}\,.
\end{equation}
Here, the coordinates $x$ and $y$ denote the test particle position in the plane of the planets, $r$ is its heliocentric distance, and $K(\xi)$ is the complete elliptic integral of the first kind\footnote{To avoid ambiguity caused by the discrepancy between the notation used for function tables and numerical implementations, we give the explicit form $K(\xi) = \int_0^{\pi/2} (1 - \xi^2 \sin{\theta})^{-1/2}\, d\theta$. Note that the modulus here is different than the one used in \citet{absteg1972}.}. While there is not a general closed-form analytic solution for $\mathcal{K}$, the expression can be integrated numerically for any appropriate values of the variables $a$, $e$, $i$, and $\omega$.

\begin{figure}[h]
    \centering
    \includegraphics[width=0.95\textwidth]{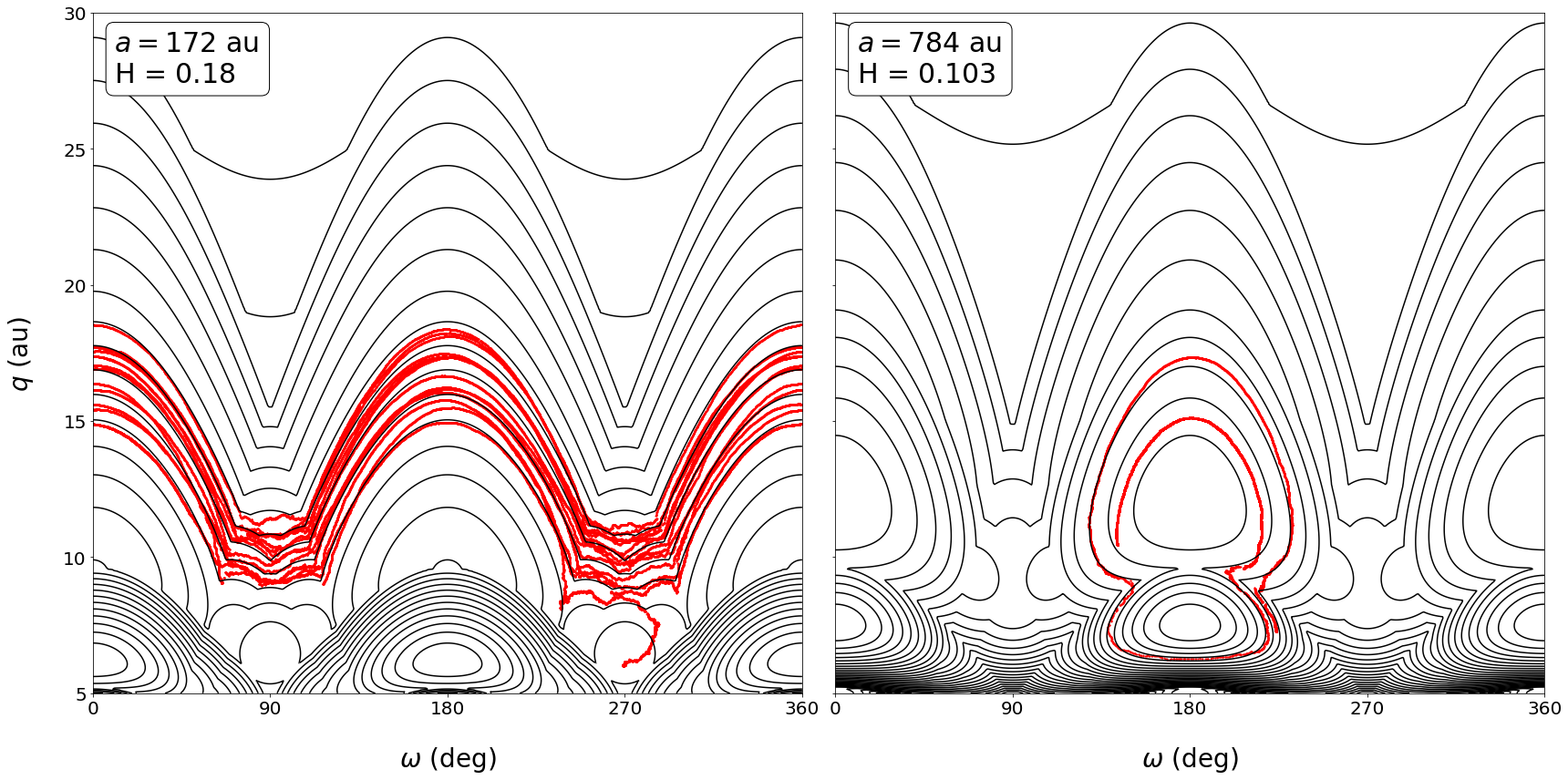}
    \caption{(Black) Hamiltonian level curves in the parameter space of $q$ vs. $\omega$ for two objects that survived for more than 500 Myr. The level curves were computed by holding constant the object's median semi-major axis $a$ and reduced action $H = \sqrt{1 - e^2}\cos{i}$, as indicated in the upper-left of each panel. (Red) Numerical data from our simulations. The theory and numerics match remarkably well for these objects, suggesting that their dynamics are well-described by a simple vZLK model.}
    \label{fig:hamiltonian}
\end{figure}

In Figure \ref{fig:hamiltonian}, we plot the level curves of $\mathcal{K}$ in the parameter space of $q$ versus $\omega$ by holding fixed $a$ and $H = \sqrt{1-e^2}\cos{i}$. In each panel, we overlay in red the simulation data of the object used to generate the level curves. The theoretical curves match the numerical data rather well, indicating that for these objects the vZLK effect provides a good working description of their long-term dynamics. We find that this theoretical framework matches the numerical data for 11 of the 13 objects that survived for more than 500 Myr. We do not expect this theory to hold for objects that are close to mean-motion resonances (in which case one would have to consider extra terms in the Hamiltonian to account for resonance), or those that experience close encounters with the giant planets (in which case their $a$ and $H$ values are not well-approximated as constant). The interpretation that emerges here is that the vZLK mechanism acts as an effective phase-protection mechanism. Objects that can attain $q \sim 10$ au execute vZLK cycles and return to pericenter only at $\omega \sim 90\degree$ or $270\degree$. This geometry implies that they come to pericenter directly above and below the plane, so they are shielded from encounters with the giant planets.

\section{Population Estimates for Captured Alien Rocks}
\label{sec:populations} 

This section utilizes the newly derived lifetime function to estimate the current population of alien rocky bodies in our solar system (i.e., rocks that have been captured from other planetary systems). We thus refine our earlier estimates from Paper I. Note that rocks can be captured by the solar system during two regimes of interest: early times when the Sun still resides in its birth cluster and later times when the Sun resides in the field. 

In Paper I, we estimated that the total mass in rocky bodies captured during the cluster phase is approximately $M_{R0}\approx10^{-3}M_\oplus$. Note that the residence time of the solar system in its birth cluster is expected to be much shorter than its current age \citep{Adams2010}. As a result, all of the rocks captured from the cluster have approximately the same chances of expulsion, so that the mass in captured rocks remaining at the present epoch is given by 
\begin{equation}
M_{RC} = M_{R0} f(\age) \approx
M_{R0} \left({\tau\over\age}\right)^{8/5} \approx 10^{-9} M_\oplus \approx 6 \times 10^{18}\,{\rm g} \,,
\label{leftover} 
\end{equation}
where $f(t)$ is the fraction of captured objects remaining as a function of time and $\age$ is the age of the solar system.\footnote{Note that this scale is $\sim1000$ times less massive than the Earth’s atmosphere.}

In addition to the contribution from the birth cluster, the solar system will steadily accumulate additional rocky bodies from the field. These newly added rocks are continually ejected, so that this population will reach a steady state. In the field, the solar system captures rocky objects at a rate given by 
\begin{equation}
\Gamma = n_R \langle \sigma v \rangle \,, 
\label{caprate} 
\end{equation}
where $n_R$ is the number density of rocky bodies in the solar neighborhood and the velocity-averaged cross section $\langle \sigma v \rangle$ is evaluated for local conditions. Over most of the age of the solar system, the parameters that define the capture rate (\ref{caprate}) are expected to be constant in time. Here, the cross section as a function of incoming speed has the form derived in Paper I, 
\begin{equation}
\sigma(v) = 
{\sigma_0\over(v/v_\sigma)^2[1+(v/v_\sigma)^2]^2} \,,
\end{equation}
where $\sigma_0\approx232,000$ AU$^2$ and $v_\sigma\approx0.42$ km/s. If we combine this cross section with a Maxwellian distribution of  velocities with dispersion $s$, we find the approximate result 
\begin{equation}
\langle \sigma v \rangle \approx 
{\sigma_0 v_\sigma \over \sqrt{2\pi}} 
\left({v_\sigma\over s}\right)^3\,,
\end{equation}
where this expression has been evaluated in the limit $s \gg v_\sigma$ (see Paper I). The velocity dispersion of stars in the solar neighborhood is $\sim40$ km/s \citep{binneytremaine}. Since the value of $s$ corresponds to the relative speed between bodies, we use $s\approx\sqrt{2}(40)\approx57$ km/s for the estimates of this paper.\footnote{For completeness, we note that the solar system's current velocity with respect to the local standard of rest, $v_\text{LSR}$, is somewhat slower than 40 km/s, and could have been even smaller in the distant past. If we use a conservative value $v_\text{LSR} = 20$ km/s, the velocity dispersion becomes $s\sim45$  km/s. Using this value of $s$ increases our estimate of the steady state mass of captured rocks in the field by a factor of $(57/45)^3 \approx 2$. This correction factor is less than other uncertainties inherent in the calculation.} 

Although the size distribution and number density of rocky material in the interstellar medium are not known, we can make the following estimate. The rocky bodies must originate from planetary systems. Moreover, each planetary system is expected to eject a given mass $m_1$, where we expect $m_1\approx$ few $\times M_\oplus$ \citep{ricelaughlin}. The mass density in rocky bodies is thus given by $\rho_R = m_1 n_\ast$, where $n_\ast$ is the number density of stars. The mass accretion rate of the solar system can be found by combining the above results: 
\begin{equation}
{\dot M} = \rho_R \langle\sigma{v}\rangle = m_1 n_\ast {\sigma_0 v_\sigma \over \sqrt{2\pi}} 
\left({v_\sigma\over s}\right)^3\,.
\end{equation}

The analysis of the previous section shows that after rocks are captured, the fraction of the objects remaining in bound orbits is a decreasing function of time $f(t)$. The steady-state mass contained in (initially alien) bound objects is thus given by \begin{equation}
M_{SS} = \int_0^{\age} {\dot M} dt f(\age-t) = 
\int_0^{\age} {{\dot M} \over 1 + (t/\tau)^\beta} dt \,,
\end{equation}
where $\age$ is the age of the solar system. Using $\beta=8/5$ and changing variables ($u=t/\tau=z^5$), the expression becomes 
\begin{equation}
M_{SS} = 5 {\dot M} \tau \int_0^Z {z^4 dz \over 1 + z^8} \to {5\pi \over 2\sqrt{2\pi}} {\cos(\pi/8)\over 2 + \sqrt{2}} m_1 n_\ast \sigma_0 v_\sigma \tau 
\left({v_\sigma\over s}\right)^3\,.
\label{eq:steadystate-maxwellian} 
\end{equation}
The final expression can be obtained by contour integration and by taking the limit $Z\to\infty$, which is valid in the regime where $\age\gg\tau$ (note that $\age/\tau\approx$ 5500, so that the correction is of order $\sim0.5\%$). If we evaluate this expression using typical parameter values for the solar neighborhood ($n_\ast=0.1$ pc$^{-3}$ and 
$s$ = 57 km/s; \citealt{binneytremaine}), we find
\begin{equation}
M_{SS} \approx 7 \times10^{-14} m_1 = 
4 \times 10^{14}\,{\rm g}\, 
\left({m_1 \over M_\oplus}\right) \,. 
\label{steadystate} 
\end{equation}
Comparing equations (\ref{leftover}) and (\ref{steadystate}) we find that the mass in rocks captured from the solar birth cluster greatly exceeds the steady state mass captured from the field.\footnote{Note that this scale is $\sim1000$ times less massive than Mount Everest.}

Note that we can use the same procedure to estimate the steady-state mass of captured dark matter particles. The velocity of dark matter particles in the Milky Way's dark matter halo is assumed to follow a Maxwellian distribution with a dispersion of $\sim 200$ km/s, so that $s_{DM} \sim 200\sqrt{2}$ km/s.\footnote{Note that this estimate may change if the Milky Way's dark matter halo is rotating with respect to the inertial frame.} Although the vast majority of the particles will be moving too fast to be captured, the low-velocity tail of the distribution provides an ample regime of parameter space for capture. In practice, the dark matter particles only interact gravitationally, and should therefore be captured onto orbits similar to those of the interstellar rocks in our simulations (both the rocks and the dark matter particles act as test particles). As a result, the formalism leading to Equation (\ref{eq:steadystate-maxwellian}) remains applicable, so that the total mass of captured dark matter particles in steady-state is given by 
\begin{equation}
M_{DM} \approx M_{SS} \times \frac{\rho_{DM}}{n_* m_1} \left( \frac{s_{rocks}}{s_{DM}} \right)^3 \approx 10^{17} \, {\rm g} \,, 
\label{eq:steadystate-dm} 
\end{equation}
where we have used $\rho_{DM}=0.43$ GeV cm$^{-3}$ for the dark matter density (e.g., \citealt{salucci2010,read2014}). It is interesting, though not consequential, that this value for dark matter capture is close to the geometric mean of the (baryonic) mass captured in the field and in the birth cluster.

Finally we consider how the steady state mass of captured objects compares to the instantaneous mass of interstellar rocks passing through our solar system without being captured. Rocks in the field have a typical hyperbolic excess velocity that is larger than the escape speed of the solar system (unless $r\lta1$ au) so this estimate can neglect gravitational focusing by the Sun. In this approximation, the mass of interstellar material passing through the solar system is given by $M_{\text{int}} = \rho_{\text{int}} V_{\text{SS}}$. Here, the mass density of interstellar objects $\rho_{\text{int}} \approx 0.1$ M$_{\earth}$ pc$^{-3}$. The effective volume of the solar system depends on the chosen outer boundary $R$, such that $V_{\text{SS}}=(4\pi/3)R^3$. Since the orbit of Neptune defines the edge of the solar system for many applications, we start by taking $R$ = 30 au, and find the total interstellar mass $M_{\text{int}} \approx 10^{16}$ g. This value is about 600 times smaller than the mass contained in the solar system's inventory of captured objects. However, the interstellar objects that remain in the solar system for an appreciable time tend to be captured with semi-major axis $a\sim1000$ au and perihelion $q\sim10$ au. Such objects spend only about 0.5\% of their time inside 30 au. As a result, the mass of captured objects within 30 au and the mass of unbound objects passing through that same volume of the solar system are roughly comparable. 

\section{Discussion and Conclusions}

In this paper we have examined the long-term stability of interstellar objects captured by our solar system. Using an ensemble of numerical simulations for 276,691 synthetic captured interstellar objects, the main result of this study is the determination of a dynamical lifetime function, i.e., the fraction of objects remaining bound as a function of time (Figure \ref{fig:lifetime}). The resulting lifetime function shows that --- after a few million years --- the system reaches a state where the survival probability approaches a power-law form such that $f(t) \sim t^{-1.6}$. The lifetime function $f(t)$ over the entire range of time can be fit with the function given in Equation (\ref{eq:lifetime}). These results are broadly consistent with a one-dimensional diffusion model \citep{yabushita}, although the latter produces a somewhat steeper power-law $f(t)\sim t^{-2}$ (see Section \ref{sec:diffusion} and Figure \ref{fig:lifetime}). 

This survival fraction $f(t)$ can be used in combination with the capture cross section (see Paper I) to estimate the number of objects of interstellar origin trapped in our solar system (Section \ref{sec:populations}). The mass in alien rocks remaining from capture events in the solar birth cluster ($M_{RC}\sim10^{-9}M_\oplus$) is much larger than the steady-state mass due to capture from the field ($M_{SS}\sim10^{-13}M_\oplus$). The estimated (steady-state) mass in dark matter particles has an intermediate value ($M_{DM}\sim10^{-11}M_\oplus$; see also \citealt{Lehmann2021}). These results should also be useful in understanding phenomena such as panspermia (see, e.g., \citealt{malosh2003}, \citealt{adams2005}). 

We also analyze the dynamical evolution of the captured objects (Section \ref{sec:long-lived}). We find that not all kinds of orbits are equally likely to survive for long times. Objects that can attain pericenter distances beyond Jupiter (and preferably beyond Saturn as well) are preferentially longer-lived. The numerical data suggest that the objects that can attain and maintain high inclinations and/or become locked into vZLK oscillations are more likely to avoid close encounters with the giant planets, and thus remain in the solar system for longer times. Nonetheless, it would be useful to have additional long-term integrations to provide better statistics. 

This work has some limitations, many of which could be improved with a considerable amount of computing. The most apparent shortcoming is that the tail of our lifetime function is rather sparsely populated. As only 13 objects survived for more than 500 Myr, and only 3 objects survived for the entire 1 Gyr integration, it is not clear whether the power-law lifetime continues indefinitely. We also reiterate that this work used the synthetic captured interstellar objects from Paper I as a starting point, so it is possible that the $v_\infty$ distribution of the captured objects has a different form than assumed herein. However, the distribution of ejection times is largely independent of $v_\infty$ (see Figure \ref{fig:vinf-lifetime}), so that this issue is higher order (see the discussion of Section \ref{sec:numlifetime}). These complications, combined with uncertainties in the amount of time our solar system spent in the birth cluster versus in the field, and the velocity dispersion of the ambient rocks in each setting, make our estimate of the steady-state population of captured interstellar objects uncertain as well. Another direction for future work is to incorporate galactic tides into the numerical simulations. The objects tend to be captured onto extremely elliptical and distant orbits, and spend much of their time beyond a few thousand au where galactic (or cluster) tides are no longer negligible.

As a final comment, we note that this study did not account for the hypothesized Planet Nine (see, e.g., \citealt{P9Rreview}). Since the capture process explored in Paper I relies on an initially hyperbolic orbit entering a planet's sphere of influence (and the Planet Nine sphere of influence is minuscule compared to the volume of its orbit), the inclusion of Planet Nine would have little effect on the capture cross section. However, since objects are typically captured with semi-major axes of order $\sim1000$ au, they spend most of their time in the trans-Neptunian and Inner Oort Cloud regions where the secular effects of Planet Nine become important. To get a sense for the role that Planet Nine plays in the orbital evolution of our captured objects, we performed a set of simulations on $\sim$ 40,000 of our captures in which we included a 10 Earth-mass body with $a = 500$ au, $e = 0.6$, and $i = 18\degree$ at a random phase. These simulations yielded rich dynamics that did not appear in the simulations including only the four known giant planets. While no trans-Neptunian objects (TNOs) were produced in our simulations with only the four known giant planets (though we emphasize again that they may have been generated if we had included galactic or cluster tides), our simulations including Planet Nine produced several such cases. Some objects displayed the well-understood behavior in which Planet Nine causes their orbital planes to tilt, sometimes even switching between prograde and retrograde. Finally, we note that some objects in these simulations were able to achieve osculating pericenter distances of up to 600 au (i.e., Planet Nine injected them into the Inner Oort Cloud). It is worth noting that this result is essentially the inverse of the process discovered by \citet{P9IOC} in which Planet Nine injects Inner Oort Cloud objects into the trans-Neptunian region of the solar system, and may warrant further investigation.

\acknowledgements

We thank two anonymous referees for providing useful feedback. This material is based upon work supported by the National Aeronautics and Space Administration under Grant No. NNX17AF21G issued through the SSO Planetary Astronomy Program and by the National Science Foundation under Grant No. AST-2009096.

\bibliographystyle{aasjournal}
\bibliography{references}

\end{document}